\documentclass[a4paper]{jpconf}

\usepackage{amsmath}
\usepackage{amssymb}
\usepackage{epsfig}
\usepackage{color}
\usepackage{graphicx}
\usepackage{dcolumn}
\usepackage{bm}
\usepackage{color}

\def\fig_width{3. in} 

\begin{document}

\title{Towards a new measurement of parity violation in dysprosium}
\author{N Leefer$^1$, L Bougas$^2$, D Antypas$^1$ and D Budker$^{1,2}$}
\address{$^1$ Helmholtz Institut Mainz, 55128 Mainz, DE}
\address{$^2$ University of California at Berkeley, Berkeley, CA USA 94720-7300}
\ead{naleefer@uni-mainz.de, bougas@berkeley.edu, dantypas@uni-mainz.de, budker@uni-mainz.de}

\begin{abstract}
The dysprosium parity violation experiment concluded nearly 17 years ago with an upper limit on weak interaction induced mixing of nearly degenerate, opposite parity states in atomic dysprosium. While that experiment was limited in sensitivity by statistics, a new apparatus constructed in the interim for radio-frequency spectroscopy is expected to provide significant improvements to the statistical sensitivity. Preliminary work from the new PV experiment in dysprosium is presented with a discussion of the current statistical sensitivity and outlook.
\end{abstract}

\section{Historical perspective}

The possibility of measuring the effects of parity violating (PV) weak neutral currents in atomic systems was first considered in 1959 by Zel'dovich~\cite{Zeldovich1959}, soon after the discovery of weak charged-current interactions in beta decay~\cite{Wu1957,Garwin1957}, but well before confirmation of neutral current interactions in 1973-74~\cite{Hasert1973}. Zel'dovich considered not only a weak interaction between electrons and nucleons that could interfere with electromagnetic interactions, leading to optical rotation of polarized light in an unpolarized medium, but also the possibility of inter-nucleon weak interactions that could give rise to a parity-violating electromagnetic moment of the nucleus known as the anapole moment.

The order-of-magnitude estimates performed by Zel'dovich predicted unobservably small experimental signatures, and progress in atomic parity violation saw few gains until the 1970s, when an enhancement of parity violating effects that scales as $Z^3$, where $Z$ is the nuclear charge, was proposed by M.-A. and C. Bouchiat~\cite{Bouchiat1974}. In addition to this enhancement mechanism, the Bouchiats also proposed a new technique where parity violating interactions are amplified through interference with an electric-field-induced mixing of opposite-parity states. This method has since been known as the Stark-interference technique.

Soon after these advances, several experiments succeeded in observing parity violating signals in atoms, including bismuth (Bi)~\cite{Barkov1978,McPherson1991}, lead (Pb)~\cite{Meekhof}, thallium (Tl)~\cite{Vetter1995}, and cesium (Cs)~\cite{Wood1997}. The most precise measurement to date is in Cs, with a statistical measurement uncertainty at the 0.35\% level~\cite{Wood1997}. This result was precise enough to also provide the first confirmation of the existence of the nuclear anapole moment, by comparing measurements on different hyperfine transitions. This measurement provides important low-energy tests of high energy theory, including bounds on the running of the weak-mixing angle at low momentum transfers~\cite{Porsev2009}, and bounds on the strength of meson coupling constants from the size of the nuclear anapole moment~\cite{Haxton2002}.

In more recent years, our group has reported on two parity violation experiments in more exotic atomic systems, ytterbium (Yb) and dysprosium (Dy). Both systems were predicted to gain enhancements both from their high nuclear charge and the existence of closely spaced opposite parity levels that enhances weak mixing effects. In the case of dysprosium, a precise measurement found no evidence for parity violating effects, limited by the statistical uncertainty of the experiment. The Yb experiment was ultimately successful at measuring the largest PV amplitude yet observed in an atomic system, approximately two-orders of magnitude larger than in Cs, reporting a PV-induced amplitude with a precision of 9\% that is consistent with theoretical predictions~\cite{Tsigutkin2009}.

Both Dy and Yb are considerably more complicated to deal with from a theory standpoint, however both offer the benefit of chains of seven naturally abundant isotopes. The comparison of PV effects in different isotopes allows some of the theory uncertainty to be removed, and can provide information on an important nuclear structure question known as the `neutron-skin' problem; what is the distribution of weak-nuclear charge (carried by neutrons) relative to the nuclear charge (carried by protons)~\cite{Brown2009}. Both elements also contain isotopes with and without nuclear spin, allowing for the possibility of measuring anapole moments in both elements as an important clarification of the anapole moment disagreement between the Cs and Tl results (only an upper limit was found for Tl~\cite{Ginges2004}).

For a review of the Yb result we refer the reader to Refs.~\cite{Tsigutkin2009,Tsigutkin2010}. References~\cite{Nguyen1997,Nguyen1999} describe measurements of PV in Dy with pulsed laser excitation. The final result for this experiment was a measured value of the weak matrix element, $H_w$, $ |H_w/2\pi| = |2.3\,\pm\,2.9(\mathrm{statistical})\,\pm\,0.7(\mathrm{systematic})|$~Hz. This was in contrast to a predicted value of $H_w/2\pi = 70\pm40$~Hz at the time~\cite{Dzuba1994}. A more recent prediction based on a refined calculation places the value at $H_w/2\pi = 4 \pm 4$~Hz~\cite{Dzuba2010}. A new dysprosium apparatus was constructed in the interim for radio-frequency spectroscopy of dysprosium, leading to constraints on a wide range of physics beyond the standard model~\cite{Leefer2013a,Hohensee2013}. With this apparatus we expect significant improvements in statistical sensitivity, and have recently revived the PV measurements. In these proceedings we provide a brief overview of the unconventional Stark-interference technique that forms the basis of the Dy PV experiment, and discuss current progress towards a new and more sensitive measurement of parity violation in Dy.

\section{Dysprosium}

Atomic dysprosium is remarkable for containing two nearly degenerate electronic states of opposite parity, labeled $A$ (even) and $B$ (odd), with an electric-dipole transition between them. Both states have total electronic angular momentum $J=10$ and are found at an energy of 19798 cm$^{-1}$ above the ground state. The energy splitting between states varies as a function of isotope, with stable isotopes at atomic masses of $A = 156,158$ and $A = 160\rightarrow164$. Two isotopes, $^{163}$Dy and $^{161}$Dy, have nonzero nuclear spin, $I = 5/2$. Typical energy intervals between various sublevels of $A$ and $B$ correspond to frequencies in the range of $-2000$~MHz to $+2000$~MHz. Positive frequencies indicate that level $B$ is higher in energy than $A$.

\begin{figure}[t]
\begin{center}
\includegraphics[width=\columnwidth]{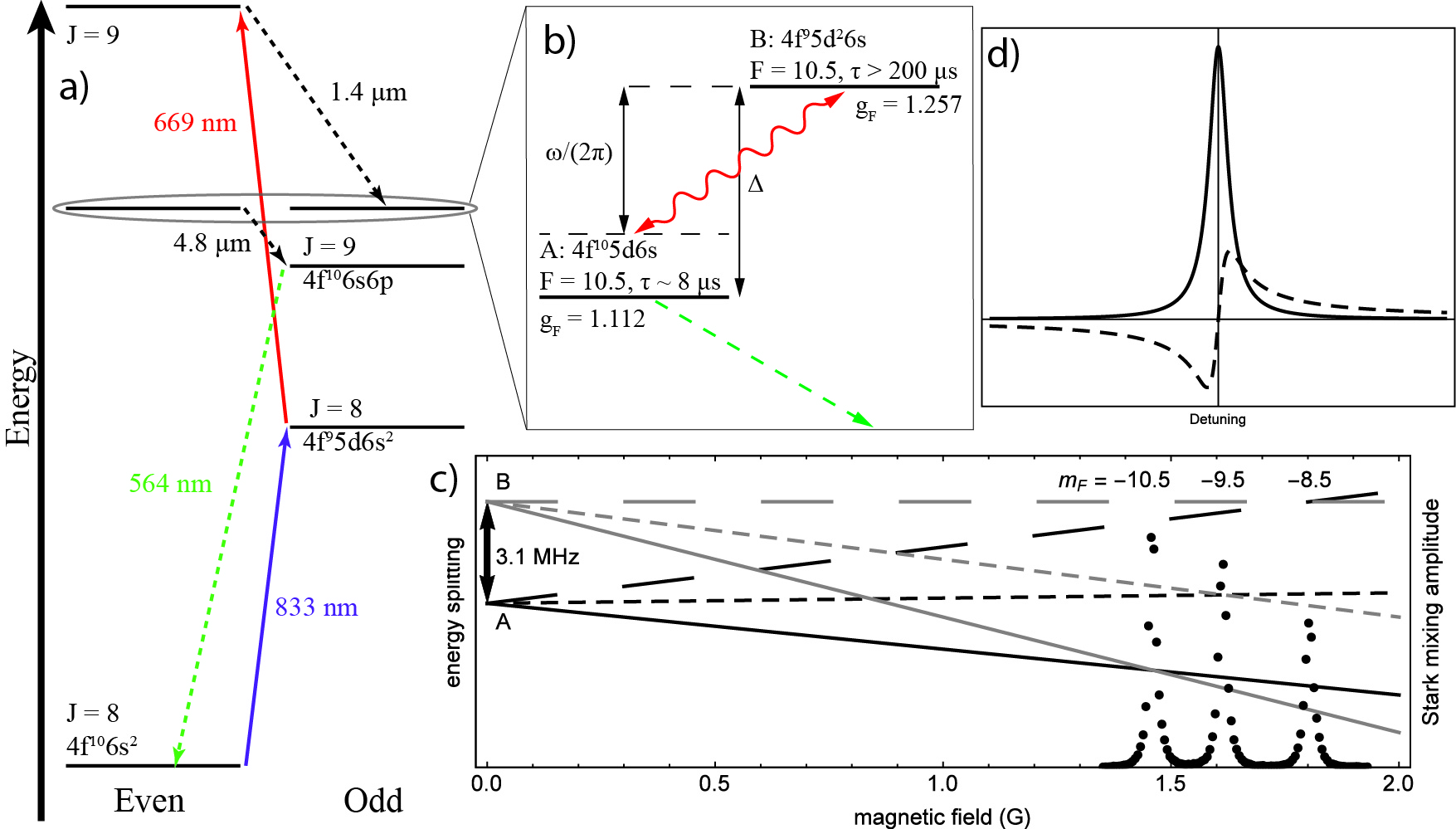}
\caption{\label{fig1}  \textbf{(a)} Energy level diagram of dysprosium showing the nearly degenerate pair of energy levels, and the transitions used for state preparation and fluorescence detection. \textbf{(b)} Zoom of Zeeman sublevels of states $A$ and $B$ brought near crossing with a magnetic field. The wavy red line indicates the applied ac electric field and the dashed straight line indicates spontaneous decay channels. \textbf{(c)} Example crossing spectrum for the first three $\Delta m_F=0$ transitions of the 3.1~MHz transition in $^{163}$Dy. The Zeeman shifts indicated by straight lines are drawn relative to the $m_{F_B}=-8.5$ sublevel for display purposes. \textbf{(d)} Calculated first-harmonic signals for a level crossing in the presence of an ac-electric field and a dc-electric field [solid line] or a PV matrix element [dashed line] for $d E_{dc} = H_w$.}
\end{center}
\end{figure}

The discussion can be general for all isotopes and hyperfine transitions in Dy, however we focus on the $F_B = 10.5\rightarrow\,F_A=10.5$ transition in $^{163}$Dy. The splitting for this transition corresponds to the remarkably small frequency of only $3.1$~MHz. Owing to a difference in g-factors between states, $g_A(F=10.5) - g_B(F=10.5) = 1.112 - 1.257$, the Zeeman sublevels for these hyperfine states can be brought to complete degeneracy with only modest magnetic fields (the $m_F = \pm10.5$ sublevels cross at $\sim1.45$~G). The level structure and diagram illustrating the level crossing is shown in Fig.~\ref{fig1}.

In the dysprosium experiment we prepare a thermal atomic beam of Dy atoms in state $B$, which is metastable, and apply an oscillating electric field to off-resonantly excite atoms to state $A$. State $A$ is short lived and decays to the ground state. One decay channel from state $A$ emits a 564-nm photon that is used for detection of the $B\rightarrow A$ transition. The geometry of the experiment is such that only $\Delta m_F = 0$ transitions are induced. The weak-induced mixing between $A$ and $B$, parameterized by an effective matrix element $H_w$, interferes with the electric field driven transition amplitude. This interference can be observed in the 564-nm fluorescence as a beat signal at the frequency as the oscillating electric field. Because magnetic sublevels of the states $A$ and $B$ can be brought to crossing, the system is accurately represented as a two-level system. The effective Hamiltonian (including non-Hermitian terms for spontaneous emission) describing the system can be written as

\begin{equation}\label{eqn1}
H = \left (\begin{array}{ccc}
    -i \Gamma_A/2 & d\,E+i\,H_w\\
    d\,E-i\,H_w & \Delta-i\,\Gamma_B/2\\
    \end{array}\right),
\end{equation}
where $\Gamma_A/2\pi \approx20$~kHz and $\Gamma_B/2\pi<1$~kHz are the natural linewidths of states A and B, $\Delta$ is the energy separation of $A$ and $B$, $i\,H_w$ is the PV matrix element, $E$ is the applied electric field, and $|d/2\pi| = 3.8(2)$~kHz/(V/cm) is the dipole matrix element between states $A$ and $B$~\cite{Budker1994}. We use the Schr\"{o}dinger equation to solve for the time-dependent wavefunction $|\psi(t)\rangle$ and compute the quantity $|\langle A|\psi(t)\rangle|^2$, which is directly proportional to the observed fluorescence in the experiment. The details of this calculation can be found in Refs.~\cite{Nguyen1997,Nguyen1999}. Assuming an electric field with both an oscillating (ac) and static (dc) component, $E = E_{ac}\cos{\omega t}+E_{dc}$, we ultimately find that

\begin{equation}
|\langle A|\psi(t)\rangle|^2 \approx\left(\frac{d E_{ac}}{\omega}\right)^2\sin^2{\omega t}+2\frac{d^2 E_{ac}E_{dc}}{\omega}\frac{\Gamma_A}{\Delta^2+\Gamma_A^2/4}\sin{\omega t} - 2 \frac{d E_{ac}H_w}{\omega}\frac{\Delta}{\Delta^2+\Gamma_A^2/4}\sin{\omega t},\label{eqn2}
\end{equation}
where the primary simplifying assumption is that $\omega$ is larger than all other frequencies in the system. 

\begin{figure}[t]
\begin{center}
\includegraphics[width=\columnwidth]{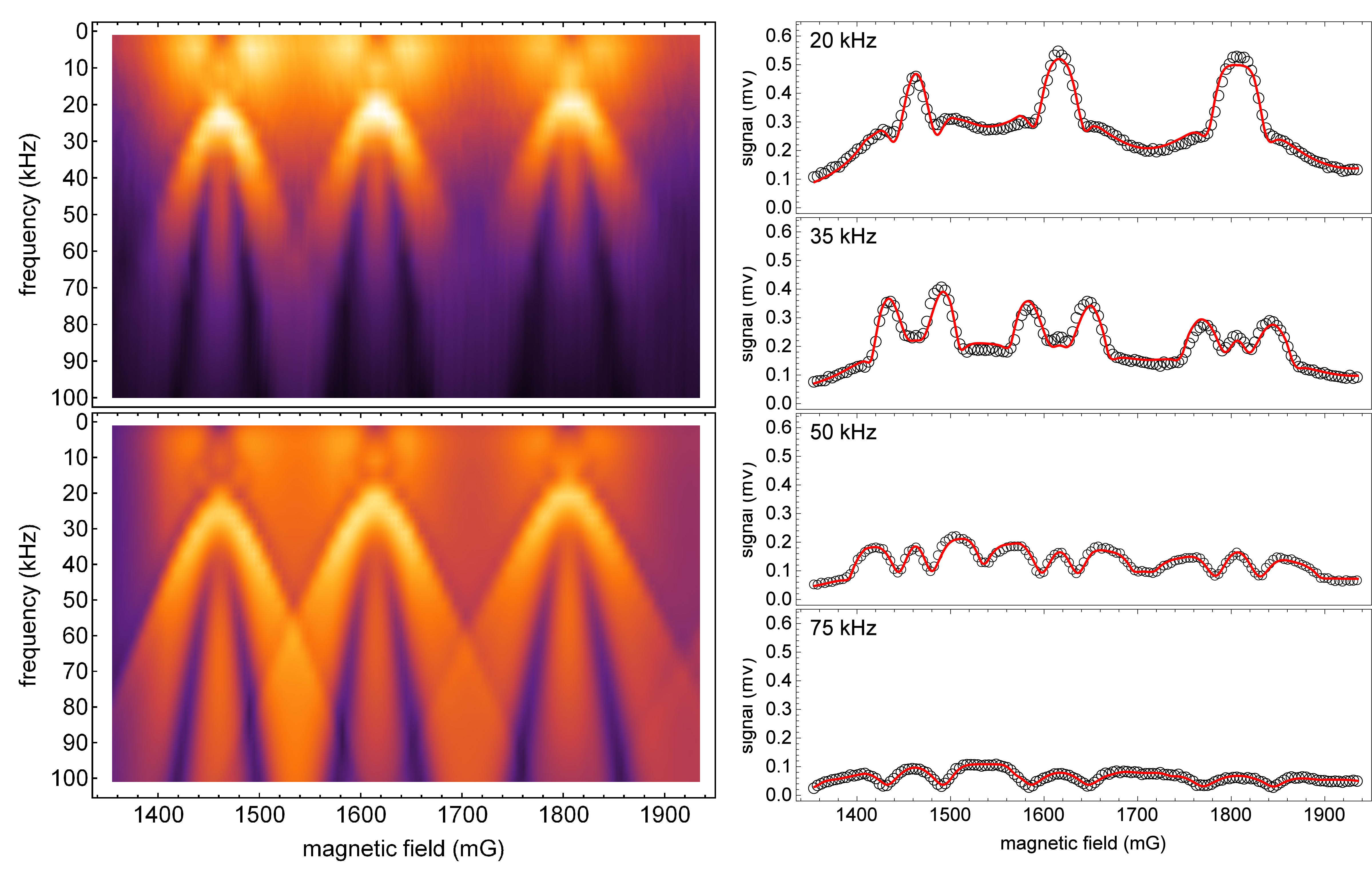}
\caption{\label{fig2} (\emph{top left}) Density plot of the observed second-harmonic amplitude as a function of magnetic field and electric-field frequency. Lighter shading corresponds to higher amplitude. Crossings of the $m_F = -10.5, -9.5,\,\mathrm{and}\,-8.5$ states in order of increasing magnetic field are easily distinguished. (\emph{bottom left}) Simulation of the expected signal for similar experimental conditions, assuming equal populations of $m_F$ sublevels. The fall-off of the observed signal with the modulation frequency is not numerically reproduced, and is still a topic of investigation. (\emph{right}) Constant frequency cuts through the experimental density plot. Transition amplitudes were saturated, so the relative signal amplitudes are determined by the sublevel populations. Solid lines are least-squares fits of constant frequency slices of the simulation to these cuts. Free parameters of the fit were an independent amplitude for each crossing and the ac-electric field amplitude. The value of the electric field obtained from the fits is consistent with the expected value within 10\%.}
\end{center}
\end{figure}

Equation~\eqref{eqn2} illustrates the primary features of the experiment. The time-dependent fluorescence signal contains an oscillating component at the second harmonic ($2\omega$) of the ac-field frequency, and components at the first-harmonic ($\omega$) in the presence of dc-fields and/or a non-zero PV matrix element. These harmonics can be acquired with standard lock-in detection techniques, and taking the ratio of the first- and second-harmonic signals allows for elimination of overall scaling factors, \textit{e.g.} light-collection efficiency and total atom number. Examination of Eq.~\eqref{eqn2} also illustrates how the PV signal can be distinguished from a dc-field. Both a dc-field and PV signal reverse sign on reversal of $E_{ac}$, however only the PV signal reverses sign on a reversal of the detuning, $\Delta$, or a reversal of the magnetic field (the matrix element $d$ reverses sign with a reversal of the Zeeman sublevel $m_F$, equivalent to reversing the magnetic field). 

An example of the expected lineshapes predicted from Eq.~\eqref{eqn2} is shown in Fig.~\ref{fig1}. Real lineshapes acquired with the current apparatus are found to be more complicated. A closed form, fully analytic solution to the general two-level problem does not exist, however a numerical integration of the Liouville equations that includes a model for lock-in detection has been found to reproduce the observed lineshapes generally well. This is demonstrated in Fig.~\ref{fig2}, which shows both real and simulated amplitudes for the second-harmonic signal as a function of magnetic field and ac-field frequency. Further refinements of these simulations are ongoing, as we do not yet understand several features, for instance, the fall-off of the signal amplitude with frequency.

\section{Experimental sensitivities and outlook}

\begin{figure}[t]
\begin{center}
\includegraphics[width=\columnwidth]{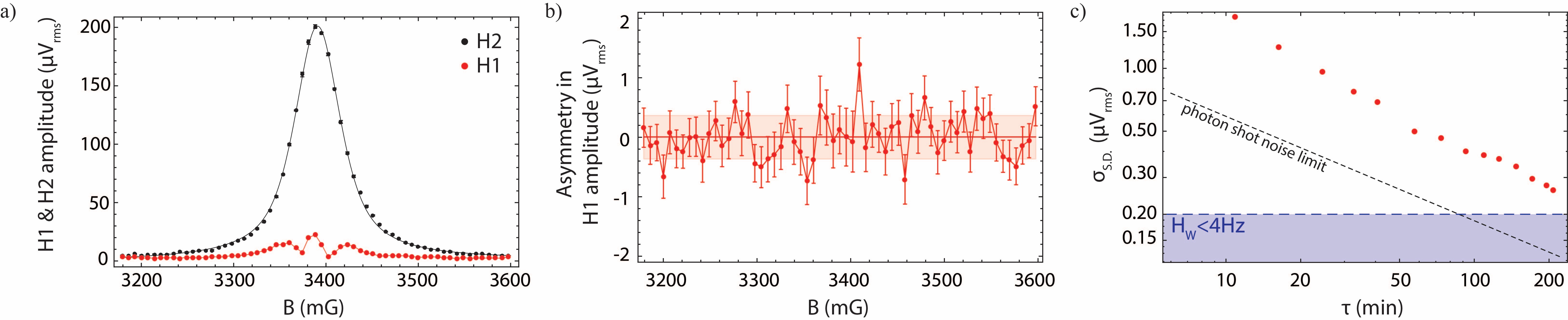}
\caption{\label{fig3} Crossing spectrum of the $|F_B=10.5,m_{F_B}=4.5\rangle\rightarrow |F_A=10.5,m_{F_A}=4.5\rangle$ states. \textbf{(a)} First and second harmonic amplitude as a function of the magnetic field for $\omega=13.1$\,kHz, $E_{ac}=9$\,V/cm, and $E_{dc}=200$\,mV/cm (see Eq.\,\ref{eqn2}). \textbf{(b)} Asymmetry in the first-harmonic signal amplitude after the application of the reversals, followed by averaging over all the acquired lineshape scans (see text for discussion). \textbf{(c)} Standard error of the averaged first harmonic signal at 3353\,mG (FWHM) where the PV effect is expected to be the largest, as a function of the acquisition time. The dashed-black line shows the projected standard error for a photon shot-noise-limited measurement, and the horizontal dashed-blue line demonstrates the sensitivity limit for achieving a $H_w/2\pi<4$\,Hz sensitivity.}
\end{center}\end{figure}

We perform studies of the Zeeman-tuned hyperfine level crossings of the $|F_B=10.5,m_{F_B}\rangle\rightarrow |F_A=10.5,m_{F_A}\rangle$ states for both signs of the electric and magnetic fields, with typical reversing rates of $\sim$1\,Hz for each field. In particular, the first- and second-harmonic amplitudes for the Zeeman-crossing spectrum of the $|F_B=10.5,m_{F_B}=4.5\rangle\rightarrow |F_A=10.5,m_{F_A}=4.5\rangle$ states are shown in Fig.\,\ref{fig3}\,a. The $m_F = 4.5$ states are chosen to work with a well isolated spectrum. We have identified stray dc electric fields that vary from 50-100\,mV/cm from day to day, and stray magnetic fields of $\sim260\,\mu$G. For each recorded point, we reverse the electric and magnetic fields, and compute the resulting asymmetries in the first harmonic signal for each profile scan, followed by averaging the results over all the scans (Fig.\,\ref{fig3}\,b). 

Analysis of our measurements suggests that it is possible to achieve the statistical sensitivity reached in the original PV measurement with only $\sim4$~hours of data, in contrast to $\sim30$~hours before~\cite{Nguyen1997}. There is still significant room for improvement, however, as we are still a factor of $\sim$2.5 above the projected shot-noise limit (see Fig.\,\ref{fig3}\,c), and are using $m_F$ states with generally lower atom numbers due to laser-population transition amplitudes. We expect up to an order-of-magnitude better statistics using 421-nm light to optically pump atoms to the maximum $m_F$ states~\cite{Leefer2010}, which will also eliminate the overlap problem. Such an improvement will be necessary for reaching our intended statistical sensitivity to weak matrix elements, $H_w/2\pi \leq 10$~mHz.\\

\noindent\textbf{Acknowledgements} NL was supported by a Marie Curie International Incoming Fellowship within the 7th European Community Framework Programme.
\\

\bibliography{PNCDy}

\end{document}